\documentclass{article}
\usepackage[utf8]{inputenc}
\usepackage{comment}
\usepackage{authblk}
\usepackage{color}
\usepackage{epsfig}
\usepackage{amsfonts}
\usepackage{floatrow}
\newfloatcommand{capbtabbox}{table}[][\FBwidth]
\usepackage{blindtext}
\usepackage{amsmath}
\usepackage{amssymb}
\usepackage{rotating}
\usepackage{cancel}
\usepackage{framed}
\usepackage{hyperref}
\usepackage{graphicx}
\usepackage{multirow}
\usepackage{chngpage}
\usepackage{floatrow}
\usepackage{blindtext}

\def\ci{\perp\!\!\!\perp}
\usepackage{amsmath, amssymb}

\title{Instrumental Variable methods to target Hypothetical Estimands with longitudinal repeated measures data: Application to the STEP 1 trial}

\author[1,2,3]{Jack Bowden}
\author[2]{Jesper Madsen}
\author[2]{Bryan Goldman}
\author[2]{Aske Thorn Iversen}
\author[3]{Xiaoran Liang}
\author[4]{Stijn Vansteelandt}

\affil[1]{Novo Nordisk Research Centre (NNRCO), Oxford, U.K}
\affil[2]{Biostatistics, Novo Nordisk, D.K}
\affil[3]{Exeter Medical School, University of Exeter, Exeter, U.K}
\affil[4]{Department of Applied Mathematics, Computer Science and Statistics, Ghent University, Belgium}
\affil[*]{Correspondence: jbwd@novonordisk.com}

\begin{document}
\maketitle
\begin{abstract}
\noindent The STEP 1 randomized trial evaluated the effect of taking semaglutide vs placebo on body weight  over a 68 week duration. As with any study evaluating an intervention delivered over a sustained period, non-adherence was observed. This was addressed in the original trial analysis within  the Estimand Framework by viewing non-adherence as an intercurrent event. The primary analysis applied a treatment policy strategy which viewed it as an aspect of the treatment regimen, and thus made no adjustment for its presence. A supplementary analysis used a hypothetical strategy, targeting an estimand that would have been realised had all participants adhered, under the assumption that no post-baseline variables confounded adherence and change in body weight. In this paper we propose an alternative Instrumental Variable method to adjust for non-adherence which does not rely on the same `unconfoundedness' assumption and is less vulnerable to positivity violations (e.g., it can give valid results even under conditions where non-adherence is guaranteed). Unlike many previous Instrumental Variable approaches, it makes full use of the repeatedly measured outcome data, and allows for a time-varying effect of treatment adherence on a participant's weight. We show that it provides a natural vehicle for defining two distinct hypothetical estimands: the treatment effect if all participants would have adhered to semaglutide, and the treatment effect if all participants would have adhered to both semaglutide and placebo. When applied to the STEP 1 study, they both suggest a sustained, slowly decaying weight loss effect of semaglutide treatment.
\end{abstract}
{\bf Key words:}  Causal Inference, Instrumental Variables, Longitudinal Data, G-estimation.

\section{Introduction}

\begin{figure}[htbp]
\centering
\includegraphics[width=.7\textwidth,clip]{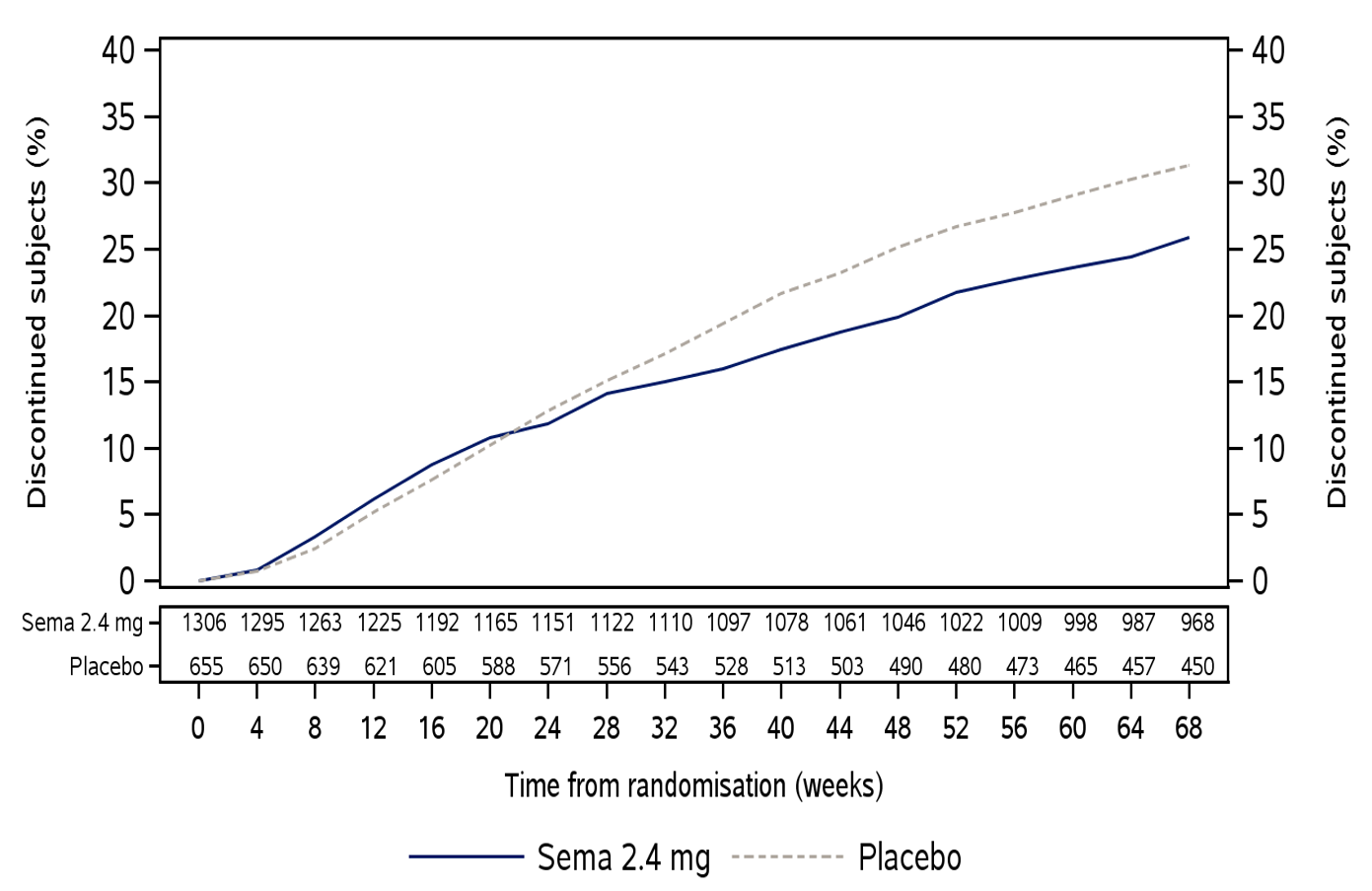}
\includegraphics[width=.75\textwidth,clip]{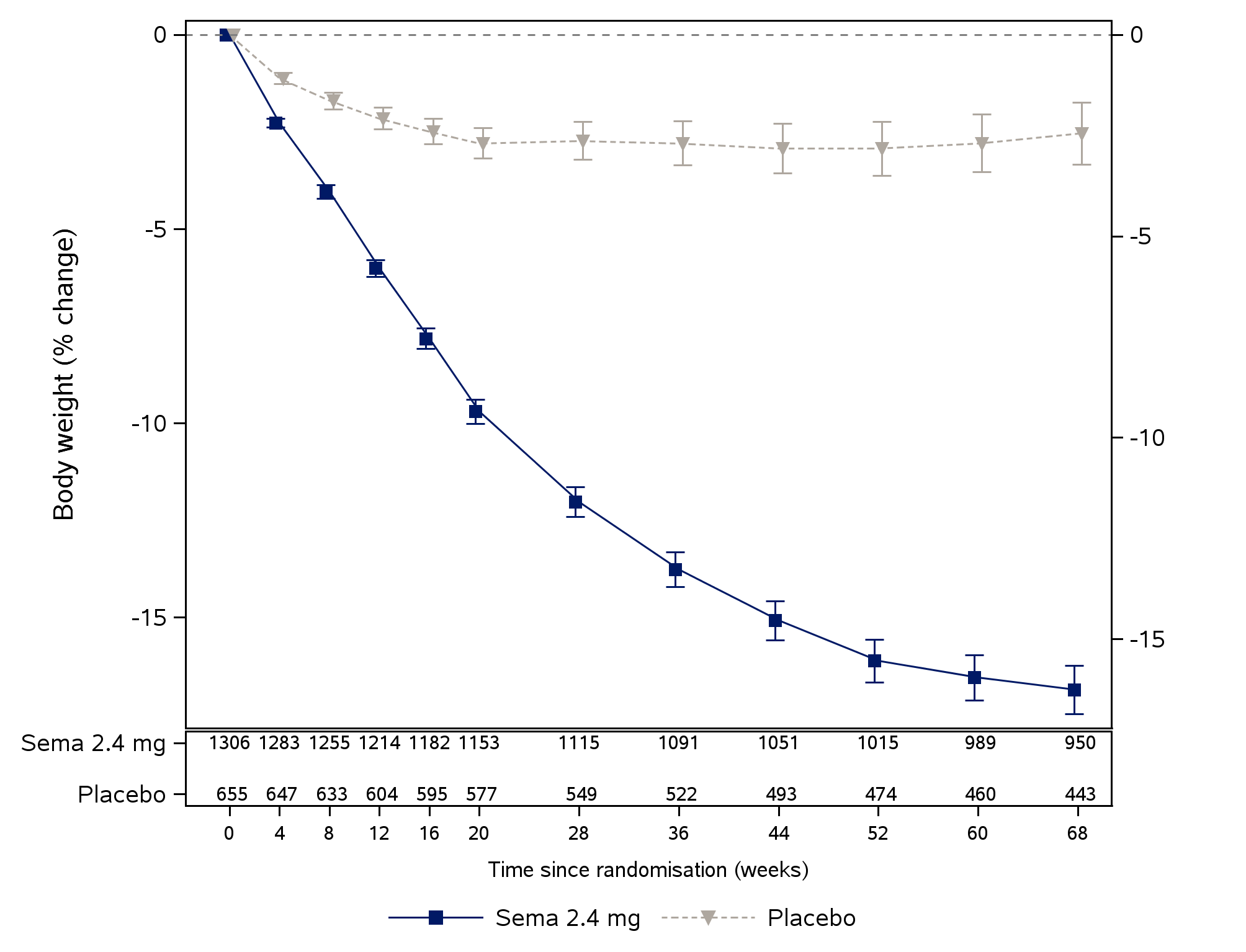}
\caption{{\it Top: Time to discontinuation of the trial product in each arm over 68 week trial duration. Bottom: Estimated weight-loss trajectories in the semaglutide and Placebo arms over the 68 week trial duration of STEP 1 under the MMRM.}}
\label{fig:STEP1}
\end{figure}

The growing prevalence of obesity is recognised as one of the most serious health issues of the 21st century. A healthy lifestyle, encompassing a balanced diet and sufficient physical exercise, is a key pillar of the weight management puzzle \cite{Curioni2005}, but many socio-economic and cultural factors, existing co-morbidities and even an individual's genetics \cite{frayling2007,patel2024} make this a much harder prospect for some people than others. Clinical guidelines suggest an increasing role for pharmacological and surgical interventions in the treatment of obesity \cite{nudel2019}.\\
\\
Semaglutide was initially developed as a treatment for people living with type II diabetes in order to improve glycaemic control, but its additional effect on body weight led to its subsequent approval for the treatment of obesity. A pivotal trial supporting its use in weight loss was STEP 1 \cite{wilding2021}. It randomised 1961 adults with a body mass index of over 30, or 27 with co-morbidities, to receive either a weekly 2.4 mg injected dose of semaglutide or a sham placebo injection, as an adjunct to lifestyle intervention. In-house participant visits occurred at twelve time points, with the primary outcome being mean change in body weight from baseline at week 68.\\
\\
As with any trial evaluating an intervention delivered over a sustained period, a degree of non-adherence was observed.  By the end of treatment at week 68, approximately 31\% of the placebo arm and 26\% of the semaglutide arm had experienced a first discontinuation episode (Figure \ref{fig:STEP1} Top), with some pausing treatment momentarily and others stopping treatment permanently. The trial analysis treated this as an intercurrent event and accounted for its presence within the Estimand Framework \cite{ICHE9}. The primary {\it Policy} strategy viewed discontinuation as an integral part of the therapy's total efficacy. The analysis used all weight loss measurements, including those taken after treatment cessation. Approximately 10\% of subjects had missing week 68 weight outcomes, which was addressed with multiple imputation, employed separately in each treatment arm via a retrieved dropout strategy. It revealed a 14.9\% reduction in the treatment arm and 2.4\% reduction in the placebo arm, the associated treatment difference being -12.4\% (95\% CI: -13.4\%,-11.5\%). A second {\it Hypothetical} or `trial product' strategy was also used that explicitly adjusted for discontinuation. It used a mixed model for repeated measurements (MMRM) that included participants up until the point of their first discontinuation, and adjusted for baseline body weight. This analysis assumed that missing outcome data was sequentially ignorable given baseline weight, previous adherence status and previous outcome measures \cite{parra2022}. This estimated a 16.9\% reduction in the treatment arm and a 2.4\% reduction in the placebo arm, the associated treatment difference being -14.4\% (95\% CI: -15.3\%,-13.5\%) (Figure \ref{fig:STEP1} bottom).\\ 
\\
\noindent Instrumental Variable (IV) methods have been historically proposed as a means to adjust for non-adherence in clinical trials \cite{cuzick1997, fischer2011, greenland2008}  and recently recommended for use within the Estimand Framework to adjust for intercurrent events \cite{bowden2021,michiels2024}. The appeal of this approach is that it can work even when unmeasured `confounding' variables simultaneously predict treatment discontinuation and the trial outcome. This is in stark contrast to methods that assume unconfoundedness, namely that participants' decision to discontinue treatment is independent of their counterfactual outcome under continuous treatment or control given a set of baseline and time-varying covariates \cite{qu2020general,lu2022}, or methods that assume sequential ignorability such as the MMRM \cite{parra2022}. Such assumptions are often unrealistic as the decision to discontinue treatment likely depends on unmeasured factors that arose during the trial. Moreover, unlike IV approaches, these methods also rely on the Positivity assumption, which states that all trial participants have a non-zero probability of experiencing the adherence profiles that form the basis of a given potential outcome contrast (e.g. full adherence to the active treatment and full adherence to placebo). Even so,  the application of IV methods within a trial analysis has so far been limited, with participants' treatment adherence generally being aggregated into a single binary variable and the effect of treatment considered only for a single outcome measured at the trial's conclusion \cite{cuzick1997, fischer2011, bowden2021}. This (often incorrect) simplification of the trial data conveniently side-steps the need to address the issue of time-varying confounding: that is, where time-varying measurements on adverse events and participant health influence their future adherence and outcome, which in turn influence adverse events and participant health at later time points.   
\noindent Using STEP 1 as an exemplar, we propose an IV method to adjust for non-adherence in a trial with longitudinal repeated measures data on adherence status and a continuous trial outcome - in our case weight loss. It allows for a time-varying causal effect of treatment adherence on a participant's weight and for the presence of unmeasured time-varying confounding. We show that it provides a natural vehicle for defining two distinct quantities: Hypothetical Estimand 1: The treatment effect that would have been observed had no semaglutide arm participants discontinued; and Hypothetical Estimand 2: the treatment effect that would have been observed if no semaglutide {\it or} placebo arm participants had discontinued. Our proposal leverages the extensive information provided by repeated measures data, incorporating plausible assumptions about the smoothness of the treatment effect over time, in order to gain insight into the effects of both treatment and placebo. When applied to STEP 1, our model highlights the sustained, slowly decaying weight loss effect of semaglutide. 

\section{Methods}

\begin{figure}[htbp]
\centering
\includegraphics[bb = 100 60 840 600, width=0.495\textwidth,clip]{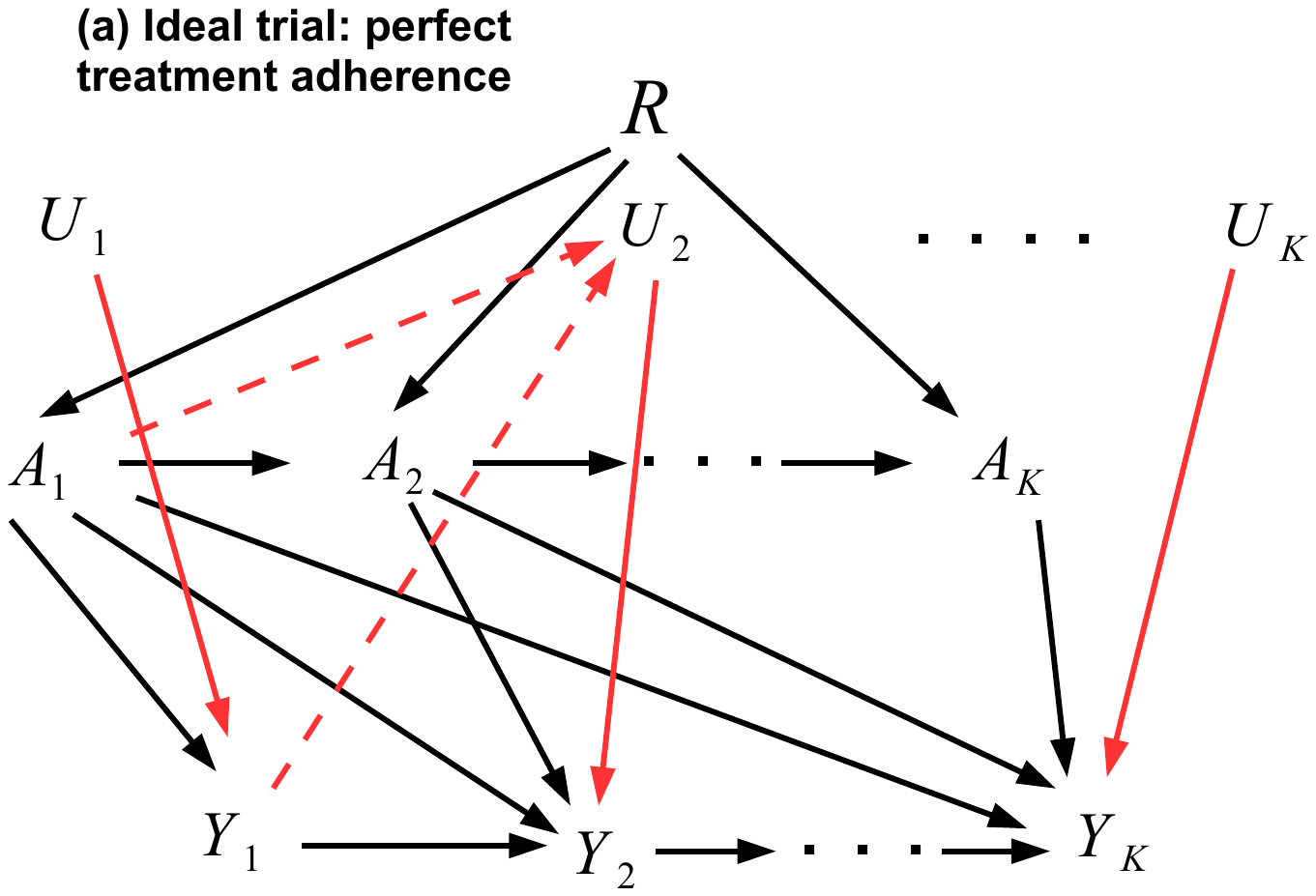}
\includegraphics[bb = 50 60 840 600, width=0.495 \textwidth,clip]{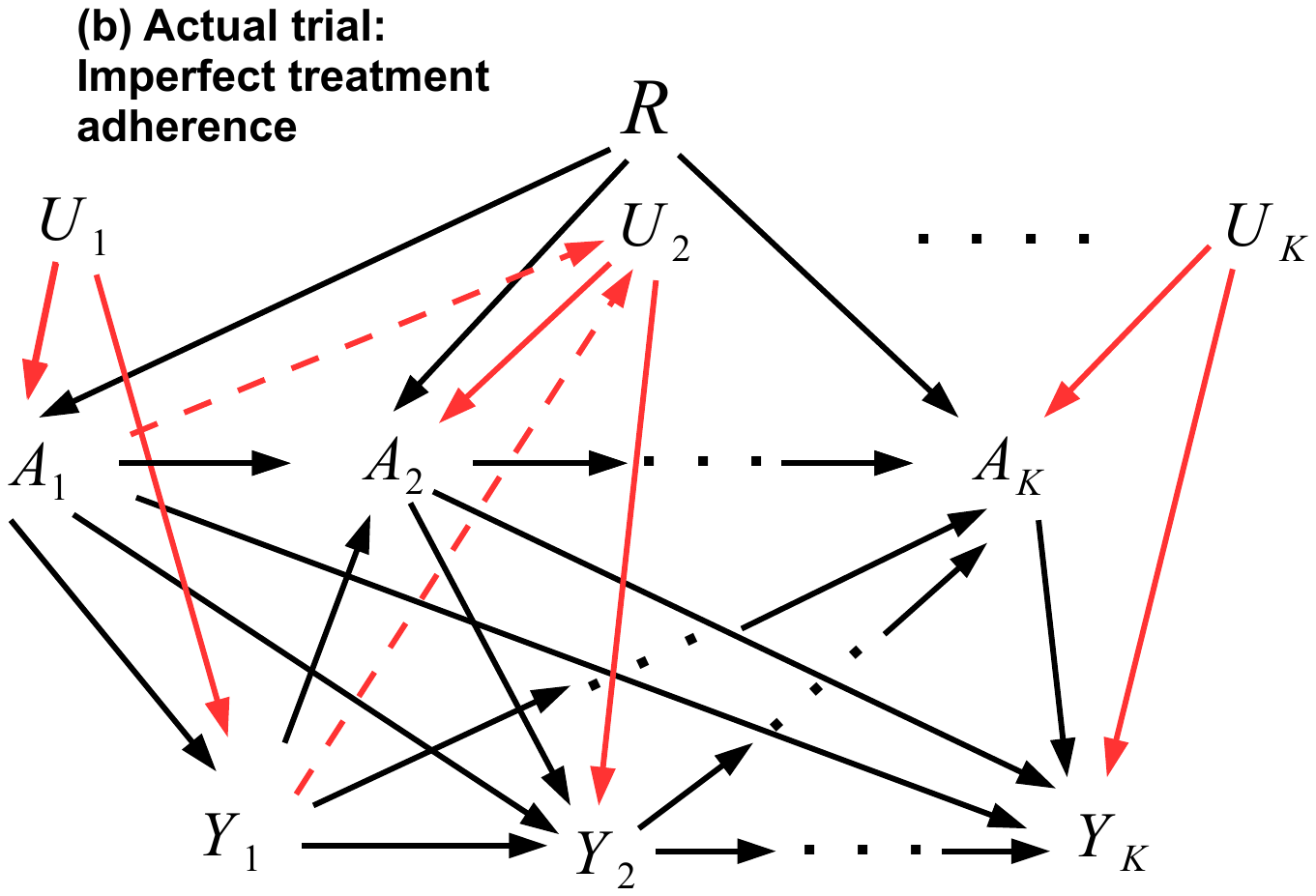}
\includegraphics[bb = 50 60 840 600, width=0.65 \textwidth,clip]{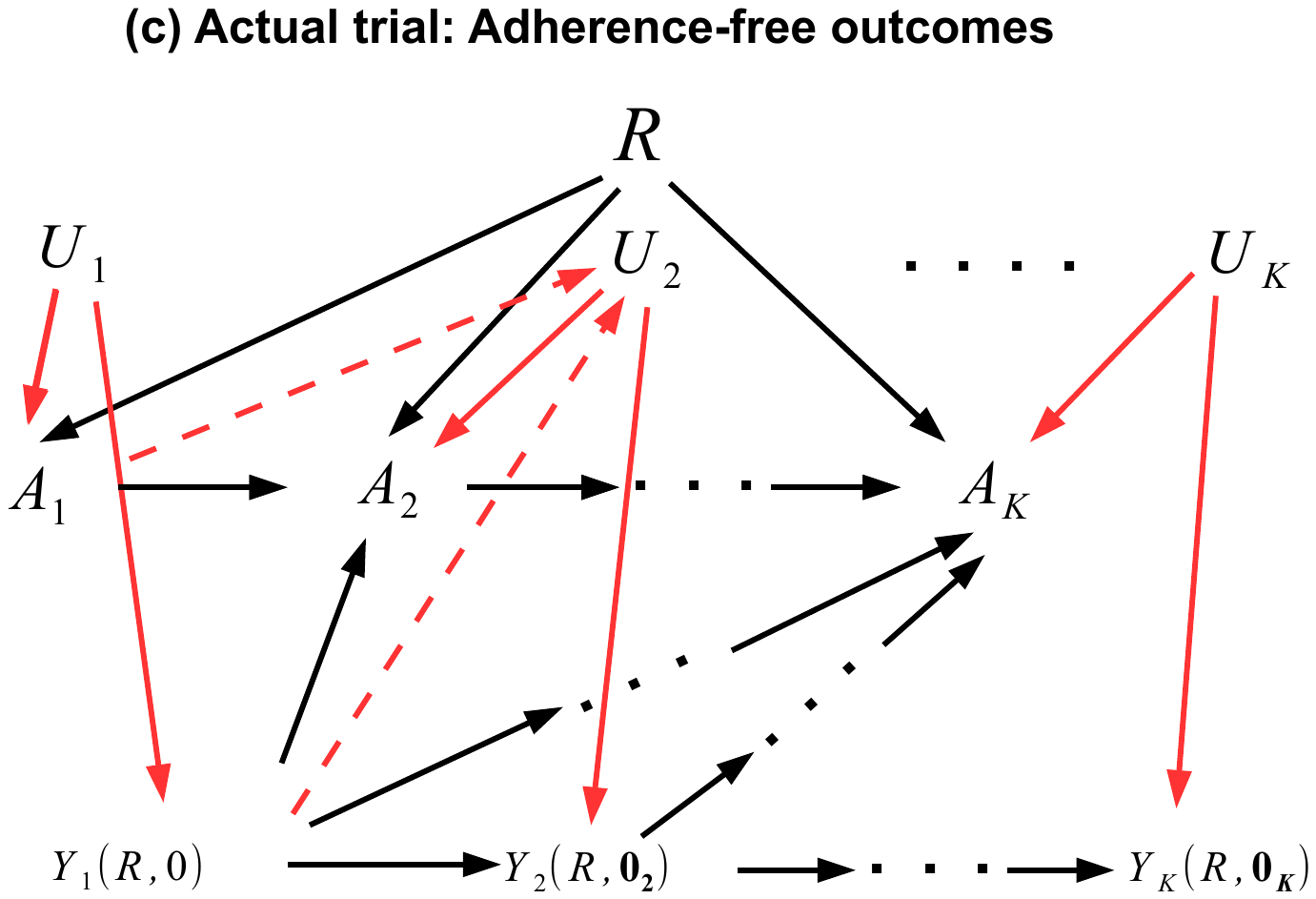}
\caption{{\it (a) Assumed causal structure for an idealised trial with perfect adherence to treatment (note no confounding of $A$ and $Y$). (b) Assumed causal structure for the actual trial with only partial adherence. Dashed red lines illustrate the possibility that a participant's earlier treatment adherence and weight may alter later treatment and weight via unmeasured confounders. (c) Approximate visualization of the implied causal structure upon replacing observed treatment arm outcomes with predicted counterfactual outcomes, which are then assumed to have the same expectation across randomized groups.}}
\label{fig:causaldag}
\end{figure}

For simplicity, we first motivate the technique of G-estimation \cite{robins1994,vansteelandt2014} by using it to adjust for non-adherence to treatment, thereby ignoring non-adherence in the placebo arm. This enables the identification and estimation of Hypothetical Estimand 1. An extended version of the model will then be introduced that adjusts for treatment and placebo arm non-adherence, which naturally leads to identification and estimation of Hypothetical Estimand 2.\\
\\
Let $R$ denote randomization to either the active treatment arm ($R=1$) - consisting of an injection of semaglutide, or the control arm ($R=0$) consisting of a sham placebo injection. Further, let $Y_{k}$ and $A_{k}$ represent, at time $t_{k}$: an individual's observed percentage weight loss and whether a participant is adherent to treatment ($A_{k}=1$) or not  ($A_{k}=0$). By design in STEP 1, placebo group participants could not access the active drug, so $A_{k}=0$ if $R=0$ for {\bf all} trial time points.\\
\\
In an idealised trial all individuals would perfectly adhere to their randomly assigned treatment so that, for all time points $k$, $A_{k}= 1$ iff $R=1$ and $A_{k}= 0$ otherwise. This is illustrated in the causal diagram of Figure \ref{fig:causaldag} (a). Note that in this case, although additional participants-specific factors, $U_{k}$, (for example side effect history, innate psychological motivation), may influence $Y_{k}$, they {\bf cannot} influence $A_{k}$ as it is uniquely predicted by $R$ in this idealised trial. This guarantees that, at the trial's conclusion, participants who are fully adherent to treatment ($A_{k}$ = 1 for all $k$) are still exchangeable with participants who were fully non-adherent to treatment ($A_{k}$ = 0 for all $k$) \cite{parra2022}. \\
\\
In reality, Figure \ref{fig:causaldag} (a) is not an accurate representation of STEP 1 because there {\it was} treatment arm non-adherence. This means that, at a given time point $k$, many different participant groups with distinct treatment adherence histories exist, as opposed to the original (intended) two groups of full treatment adherers and no-treatment adherers. This is problematic because it leaves open the possibility that an individual's:
\begin{itemize}
    \item{Adherence history $\bar{A}_{k-1}$=$(A_{1},...,A_{k-1})$;}
    \item{Weight trajectory $\bar{Y}_{k-1}$=$(Y_{1},...,Y_{k-1})$;}
    \item{Participant factor history $\bar{U}_{k}$=$(U_{1},\ldots, U_{k})$,}
\end{itemize} 
\noindent now jointly predict both $A_{k}$ {\it and} $Y_{k}$. This set up is illustrated in Figure \ref{fig:causaldag} (b). Taking, for example, data up to time point 2 in the trial and using the rules of d-separation on Figure \ref{fig:causaldag} (b), \cite{pearl2014} to preserve exchangeability between participants with any adherence history $\bar{A}_{2} \in \{(0,0),(0,1),(1,0),(1,1)\}$, we would need to adjust for $A_{1}$, $Y_{1}$ and $U_{2}$. That is, all joint predictors of $A_{2}$ and $Y_{2}$ \cite{parra2022}.  At any time-point $k$, if any components of $\bar{U}_{k}$ are either unmeasured or imprecisely measured, it is not possible to satisfy the exchangeability assumption through direct adjustment. We now describe how  G-estimation based on an Instrumental Variable (IV) - in our case the initial randomization indicator $R$ - \cite{goetghebeur1997} can be used to identify and estimate the treatment effect that would have been seen in STEP 1 if all participants had adhered to the active drug over the full 68 week duration of the trial, had they been randomised to it. Such G-estimation procedures do not require data on the unmeasured confounder $\bar{U}$.

\subsection{Randomization as an Instrumental Variable}

For $R$ to be a valid IV with respect to the treatment and outcome at time point $k$, it needs to satisfy three core assumptions \cite{bowden1990,hernan2020}. It must:
\begin{itemize}
    \item{Be a predictor - ideally a strong predictor - of $A_{k}$ (IV1);}
    \item{Not be influenced by any variables that confound the association between $A_{l}$ and $Y_{k}$ for any $l\leq k$ (IV2);}
    \item{Only influence $Y_{k}$ through $\bar{A}_{k}$ (IV3).}
    \end{itemize} 
Assumption IV1 can be empirically tested in the data. It certainly holds in STEP 1 because, at the 12th and final time point, $Pr(\bar{A}_{12}=\boldsymbol{1}_{1\times 12}|R=0)$=0 and $Pr(\bar{A}_{12}=\boldsymbol{1}_{1\times 12}|R=1) \approx 0.74$. Assumption IV2 is automatically satisfied if randomization was appropriately conducted so that a participant's characteristics did not influence their assignment to a given trial arm. This can be partially evaluated by looking at the similarity of aggregated baseline characteristics across trial arms, which were deemed to be satisfied in STEP 1 (see Table 1 in \cite{wilding2021}). Assumption IV3 will generally hold if participant care was the same across trial arms, apart from treatment administration, but its validity is heavily dependent on the definition of adherence used in the analysis, as will be discussed below.\\
\\
Provided $R$ satisfies assumptions IV1-IV3, it can be used to adjust for non-adherence in the following way. Taking ($A_{1},Y_{1}$), the adherence and outcome pair at time $t_{1}$ as an example, we can see from Figure \ref{fig:causaldag}(b) that their association is confounded by an unobserved variable, $U_{1}$. At time point 1, we can in principle observe the following three potential outcomes $Y_{1}(r,a_{1})$ for each individual:

\begin{enumerate}
\item{$Y_{1}(1,1)$: The potential outcome at time point 1 if they had been randomized to treatment and adhered};
    \item{$Y_{1}(1,0)$: The potential outcome at time point 1 if they had been randomized to treatment, but not adhered};
     \item{$Y_{1}(0,0)$: The potential outcome at time point 1 if they had been randomized to placebo and therefore not adhered to treatment (by design)};
\end{enumerate}
In a general context, and following the terminology in \cite{hernan2020}, IV2 implies that an individual's potential outcomes are independent of the group they were randomized to, or that $Y_{1}(r,a_{1}) \ci R$. Likewise, IV3 implies that $Y_{1}(r,a_{1})$ = $Y_{1}(r',a_{1})$ for all distinct pairs $r$,$r'$. In this analysis context $Y_{1}(0,1)$ is not possible, so IV3 simplifies to $Y_{1}(1,0) = Y_{1}(0,0)$. In fact, we only require the weaker condition that $Y_{1}(1,0)$ and $Y_{1}(0,0)$ are {\it distributionally equivalent}.\\
\\
Suppose that we were able to predict $Y_{1}(1,0)$ for those whom we observe $Y_{1}(1,1)$, namely by removing from $Y_1$ the unknown effect of $A_{1}$ on $Y_{1}$. This is illustrated in Figure \ref{fig:causaldag} (c) as the removal of the direct arrow from $A_{1}$ to $Y_{1}$, whose value is denoted as $Y_{1}(R,0)$. This implies that there is now no `open' path from $R$ to $Y(R,0)$, which in turns means that $R$ and  $Y(R,0)$ are independent, or `d-separated' \cite{pearl2014}. Then since the expected value of $Y(1,0)$ should be the same as $Y_{1}(0,0)$, the magnitude of this unknown effect can be chosen to force this desired equality under a parametric causal structural model for the effect of $A_{1}$ on $Y_{1}$. This is the principle of G-estimation. 
For further reading on the technique's foundation and its application in clinical trials, see \cite{fischer2011,goetghebeur1997,vansteelandt2003,bowden2011}.


\subsection{Adjusting for the time-varying effect of non-adherence in the treatment arm}\label{subsec:adjtrt}
More concretely, we now define the causal effect of $A_{1}$ within a structural mean model that links $Y_{1}(1,A_{1})$ to the (possibly) counterfactual predicted outcome $Y_{1}(1,0)$ as: 
\begin{equation}
 E\left\{Y_{1}|R=1,A_{1}\right\}-E\left\{Y_{1}(1,0)|R=1,A_{1}\right\} = \beta_{1}(1)A_{1}.
\end{equation}
\noindent 
Here, $\beta_{1}(1)$ represents the effect of being adherent to treatment until time point 1 on weight at time point 1 and we say `possibly counterfactual' because $Y_{1}(1,0)$ is directly observed for a proportion of the treatment arm.  In a similar fashion, assume that the $K-1$ potential outcomes for the remaining time points $2,...,K$ relate to the corresponding observed outcomes $Y_{2},...,Y_{K}$ via the parametric models:
\begin{eqnarray}
E\left\{Y_{2}|R=1,\bar{A}_{2}\right\}-E\left\{Y_{2}(1,\boldsymbol{0}_{2})|R=1,\bar{A}_{2}\right\}  &=&  \beta_{2}(1)A_{1} + \beta_{2}(2)A_{2}  \nonumber \\
&\vdots& \nonumber \\
E\left\{Y_{K}|R=1,\bar{A}_{K}\right\}-E\left\{Y_{K}(1,\boldsymbol{0}_{K})|R=1,\bar{A}_{K}\right\} &=& \sum^{K}_{j=1}\beta_{K}(j)A_{j},   \nonumber
\end{eqnarray}
\noindent Here,  $\beta_{k}(j)$ represents the effect of being adherent to treatment at the $j$th time point on the outcome at time $k$ and, in a simplification of notation, it is understood that  $\boldsymbol{0}_{k}$ denotes  a $k$-length vector of zeros. Assumptions IV2 and IV3 imply
\begin{equation}
E\left\{Y_{k}(1,\boldsymbol{0}_{k})|R=1\right\} = E\left\{Y_{k}|R=0\right\} \label{eq:IV23}   
\end{equation}
for $k = 1,\ldots,K$.

\subsubsection{Why is IV1 needed?}

If only outcome data at time point 1 were available, we would estimate $\beta_{1}(1)$ as the value which forced the mean treatment-free outcomes to be equal across arms, as implied by IV2 and IV3. To facilitate this, note that the estimated coefficient for $R$ in a linear regression of the combined treatment free outcomes on $R$ would then be zero. This can only occur when their covariance is zero. The treatment free outcome is $Y_{1}-\beta_{1}(1)A_{1}$ for each individual, and therefore:

\begin{eqnarray}
Cov(R,Y_{1}-\beta_{1}(1)A_{1}) = 0 &\Rightarrow& Cov(R,Y_{1}) = \beta_{1}(1)Cov(R,A_{1}) \nonumber \\
&\Rightarrow& \beta_{1}(1) = \frac{Cov(R,Y_{1})}{Cov(R,A_{1})} \label{eq:simpleIV}
\end{eqnarray}

\noindent From this we see that assumption IV1 is needed for identification by guaranteeing that the denominator of (\ref{eq:simpleIV}) is non-zero. The task of jointly estimating the full causal parameter vector $(\beta_{1}(1),...,\beta_{K}(K)$) is more complex, and will be explained in detail below, but nevertheless requires assumption IV1 for the same underlying reason. 

\subsubsection{Hypothetical Estimand 1}

An obvious Hypothetical estimand can be constructed from the parameters of the structural mean model as the contrast between the expected potential outcome under full treatment adherence up to time $K$, ($\bar{A}_{K}=\boldsymbol{1}_{K}$) versus zero treatment adherence, ($\bar{A}_{K}=\boldsymbol{0}_{K}$):
\begin{eqnarray}
\text{Hypothetical Estimand 1:} &=&E\left(Y_{K}(1,\boldsymbol{1}_{K})-Y_{K}(1,\boldsymbol{0}_{K})\right)\nonumber \\ (\text{By IV3 })&=&E\left(Y_{K}(1,\boldsymbol{1}_{K})-Y_{K}(0,\boldsymbol{0}_{K})\right)\nonumber \\
&=&\sum^{K}_{j=1}\beta_{K}(j) \label{eq:HypEst1}
\end{eqnarray}
Equation (\ref{eq:HypEst1}) requires the additional assumption that the effect of full adherence in participants who adhere up to time $K$ is the same as in other participants:
{\small \begin{equation}
E\left\{Y_{K}(1,\boldsymbol{1}_{K})-Y_{K}(1,\boldsymbol{0}_{K})|\bar{A}_{K}=\boldsymbol{1}_{K},R=1\right\}
=E\left\{Y_{K}(1,\boldsymbol{1}_{K})-Y_{K}(1,\boldsymbol{0}_{K})|\bar{A}_{K}\ne\boldsymbol{1}_{K},R=1\right\}.
\label{eq:homogeneityT}
\end{equation}}
Assumption (\ref{eq:homogeneityT}) is untestable and may well be violated because, for instance, participants who adhere to treatment up to time $K$ may possibly also be more motivated to continue adopting a healthy lifestyle, which might result in a smaller additional effect of treatment. 

\subsection{Point estimation for Hypothetical Estimand 1 parameters}
\label{sec:EstSMM1}

Each of the $K$ independence conditions enables the identification of a single causal parameter, but the full causal parameter vector $\theta$=$(\beta_{1}(1),...,\beta_{K}(K)$) has length $K(K+1)/2$ and therefore cannot be identified. To address this, we propose the following (partially untestable) two-parameter model that assumes the effect of $A_{k}$ on $Y_{k}$ is constant across all times points $k=1,...,K$ (and equal to $\beta$), but its effect thereafter follows the relation:

\begin{equation}
\beta_{k}(j) = \beta\alpha^{t_{k}-t_{j}}. \label{eq:decaymodel1}
\end{equation}

\noindent Here, $t_{k}$ represents the time that outcome $Y_{k}$ was measured and $t_{k}-t_{j}$ reflects the difference between the earlier treatment time $t_{j}$ and the later treatment time $t_{k}$. 
If $\alpha$ is less than 1 or greater than 1 this is consistent with the effect of treatment decaying over time or increasing over time, respectively, with the former being more plausible. G-estimation of $\theta$ proceeds by minimising a function of the $K$-dimensional score equation $S(\theta)$ with respect to $\theta = (\beta,\alpha)$ over individuals $i=1,...,n$, where:
\begin{eqnarray}
S(\theta) &=& \sum^{n}_{i=1}(R_{i}-\bar{R})\boldsymbol{\Sigma^{-1}}\begin{pmatrix}
Y_{1i}&-&\beta_{1}(1)A_{1i}R_{i}\\
Y_{2i}&-&\left\{\beta_{2}(1)A_{1i}+\beta_{2}(2)A_{2i}\right\}R_{i}\\
&&.\\
&&.\\
&&.\\
Y_{Ki}&-&\sum^{K}_{j=1}\beta_{K}(j)A_{ji}R_{i} 
\end{pmatrix} \nonumber \\
\label{eq:ScoreNaive}
\end{eqnarray}
\noindent Here,  $\bar{R}$ is the mean value of $R$ across all participants (e.g. $\approx \frac{1}{2}$ in the case of 1:1 randomization ) and $\boldsymbol{\Sigma}$ is an arbitrary $K\times K$ matrix. For simplicity we use the identity matrix for $\boldsymbol{\Sigma}$ throughout this paper, and opt for a minimisation function $S^T(\theta)S(\theta)$ - i.e. the sum of squared contributions of each individual score equation. Estimation of parameter uncertainty and confidence intervals for Hypothetical Estimand 1 will be discussed in Section \ref{sec:technical}.

\subsection{Accounting for non-adherence in the treatment and placebo arms}

A key assumption leveraged by the structural mean model in the previous section, where adherence is defined only with respect to the active treatment, is that counterfactual outcomes $Y_{k}(1,\boldsymbol{0}_{k})$ and $Y_{k}(0,\boldsymbol{0}_{k})$ should be equal in expectation. However, this may very well be violated. For example, subjects who discontinue early under randomization to treatment might perform worse on average than participants who are randomized and adhere to placebo. Furthermore, participants in the placebo arm lost an average of 2.4\% of body-weight in the 68 week trial, indicating a potential placebo effect of the sham injection in helping participants to adhere to the suggested lifestyle intervention. If true, then it is reasonable to assume that treatment arm participants could also be benefiting from a similar effect until the point they discontinue, which we may wish to separate out from the pure treatment effect in the analysis. Since 31\% of participants in the placebo arm were non-adherent at some point in the trial, and with the reasons for their non-adherence being potentially confounded with their weight loss, our second G-estimation strategy attempts to equate potential outcomes across trial arms upon: (a) removal of the treatment (plus any hidden placebo) effect for those in the treatment arm; and (b) removal of the placebo effect from those in the placebo arm. This provides a means for contrasting the effect of full treatment adherence with the effect of full placebo adherence.\\
\\
To this end, we now update our definition of adherence so that, at time $k$, 
\[ A_{k} = \left\{
    \begin{array}{ll}
        1 & \text{if $R$ = 0/1 and a participant is adherent to placebo/treatment}  \\
        0 & \text{if $R$ = 0/1 and a participant is non-adherent to placebo/treatment}
    \end{array}
\right. \]

\noindent Assumption IV1 holds in STEP 1 because, from Figure \ref{fig:STEP1}, it predicts adherence to treatment and placebo across all time points. At the final time point $Pr(\bar{A}_{K}=\boldsymbol{1}_{K}|R=0) \approx 0.69$ and, as previously stated, $Pr(\bar{A}_{K}=\boldsymbol{1}_{K}|R=1) \approx 0.74$. Assumptions IV2 and IV3 are defined as before, but (we argue below) are more justified with our updated definition of adherence. At time point 1, we can in principle observe the following four potential outcomes $Y_{1}(r,a_{1})$ for each individual:

\begin{enumerate}
\item{$Y_{1}(1,1)$: The potential outcome for an individual at time point 1 if they had been randomized to treatment and adhered};
    \item{$Y_{1}(1,0)$: The potential outcome for an individual at time point 1 if they had been randomized to treatment, but not adhered};
     \item{$Y_{1}(0,1)$: The potential outcome for an individual at time point 1 if they had been randomized to placebo and adhered};
     \item{$Y_{1}(0,0)$: The potential outcome for an individual at time point 1 if they had been randomized to placebo and not adhered.}
\end{enumerate}

\noindent As before, IV2 implies that, for each individual, all four potential outcomes should be independent of the trial arm they were randomized to. IV3 implies that  $Y_{1}(1,0)$ = $Y_{1}(0,0)$ and is arguably much more plausible. However, whilst both $Y_{1}(1,1)$ and $Y_{1}(0,1)$ are now observable potential outcomes, IV3 does not additionally imply $Y_{1}(1,1)$ = $Y_{1}(0,1)$, since the adherence variable now refers to two different treatments (semaglutide or placebo). Nevertheless, IV2 and IV3  imply $K$ independence conditions that can be exploited for the purposes of estimation. Namely, after a suitable transformation of participants' observed outcomes to adherence-free outcomes on {\bf both} trial arms:
\begin{equation}
E\left\{Y_{k}(1,\boldsymbol{0}_{k})|R=1\right\}=E\left\{Y_{k}(0,\boldsymbol{0}_{k})|R=0\right\}, \text{  for $k=1,...,K$.}  \label{eq:IndCond2}  
\end{equation}
\noindent We now define, for the treatment and placebo arm separately, linear structural mean models for the difference between an individual's observed outcome given their treatment and placebo adherence history at time $k$, and the value it would have taken had these adherence levels been set to 0, as: 
\begin{eqnarray}
E\left\{Y_{k}|R=1,\bar{A}_{k}\right\}-E\left\{Y_{k}(1,\boldsymbol{0}_{k})|R=1,\bar{A}_{k}\right\} &=& \sum^{k}_{j=1}\delta_{k}(j)A_{j}, \label{eq:smm1}
 \\
E\left\{Y_{k}|R=0,\bar{A}_{k}\right\}-E\left\{Y_{k}(0,\boldsymbol{0}_{k})|R=0,\bar{A}_{k}\right\} &=& \sum^{k}_{j=1}\gamma_{k}(j)A_{j}. \label{eq:smm2}
\end{eqnarray}
\noindent We parameterise the treatment arm potential outcome with a new term $\delta_{k}(j)$ to reflect the fact that it is now assumed to encapsulate the true treatment effect {\it and} the additional effect of the placebo at time point $j$ on weight at time point $k$. Likewise, $\gamma_{j}(k)$ captures only the placebo effect in the full absence of treatment. 

\subsubsection{Hypothetical Estimand 2}

To isolate the pure treatment effect at the end of the trial, we firstly define the expected difference between observed outcomes under (a)  full treatment adherence and no treatment adherence and (b) full placebo adherence and no placebo adherence. The difference between (a) and (b) then forms Hypothetical Estimand 2:
\begin{eqnarray}
\text{Hypothetical Estimand 2: }&=& 
E\left\{Y_{K}(1,\boldsymbol{1}_{K}) - Y_{K}(1,\boldsymbol{0}_{K})\right\} \nonumber\\
&-&E\left\{Y_{K}(0,\boldsymbol{1}_{K})-Y_{K}(0,\boldsymbol{0}_{K})\right\}\nonumber\\
(\text{By IV3)} &=&E\left\{Y_{K}(1,\boldsymbol{1}_{K}) - Y_{K}(0,\boldsymbol{1}_{K})\right\} \label{eq:HypEst2} \\
&=&\sum^{K}_{j=1}\delta_{K}(j)-\gamma_{K}(j),\nonumber
\end{eqnarray}
\noindent Furthermore, the last identity relies on assumption (\ref{eq:homogeneityT}) as well as 
{\small \begin{equation}
E\left\{Y_{K}(0,\boldsymbol{1}_{K})-Y_{K}(0,\boldsymbol{0}_{K})|R=0,\bar{A}_{K}=\boldsymbol{1}_{K}\right\}
=E\left\{Y_{K}(0,\boldsymbol{1}_{K})-Y_{K}(0,\boldsymbol{0}_{K})|R=0,\bar{A}_{K}\ne\boldsymbol{1}_{K}\right\}.
\label{eq:homogeneityP}
\end{equation}}
\noindent Equation (\ref{eq:HypEst2}) is comparable to the "usual" Hypothetical Estimand in the ICH E9R1 addendum \cite{ICHE9}, where treatment non-adherence is the only intercurrent event and a hypothetical strategy for this event is considered. In the discussion, we will relate assumptions (\ref{eq:homogeneityT}) and (\ref{eq:homogeneityP}) to the Principal Stratum approach of Qu et al. (2020)\cite{qu2020general} that attempts to estimate the effect in only the subset of participants who would fully adhere under assignment to either treatment or placebo, and the sequential ignorability assumption of Olarte Parra \cite{parra2022}.

\subsection{Point estimation for Hypothetical Estimand 2 parameters}
\label{sec:EstSMM2}

\noindent For this extended setting, we propose a three-parameter version of model (\ref{eq:decaymodel1}), where: 

\begin{equation}
\delta_{k}(j) = \beta\alpha^{t_{k}-t_{j}}, \quad \gamma_{k}(j) = \begin{cases}
\gamma & \text{if } j = k, \\
0 & \text{otherwise}.
\end{cases} \label{eq:decaymodel2} 
\end{equation}
\noindent We therefore assume, as before, that the effect of treatment adherence may be long-lasting, but decays over time, and impose the condition that placebo adherence only has an instantaneous effect at the time it is taken. While the  treatment and placebo arm parameters cannot be simultaneously identified using outcome data at a single time, we leverage the rich availability of repeated outcome measures to estimate the parameters by minimising $S^T(\theta)S(\theta)$ with respect to $\theta$ = ($\beta,\alpha,\gamma$) for an extended score equation based on the joint independence conditions in (\ref{eq:smm1}) and (\ref{eq:smm2}) over all time points:
\begin{eqnarray}
S(\theta) &=& \sum^{n}_{i=1}(R_{i}-\bar{R})\boldsymbol{\Sigma^{-1}}\begin{pmatrix}
Y_{1i}&-&\delta_{1}(1)A_{1i}R_{i}-\gamma A_{1i}(1-R_{i})\\
Y_{2i}&-&\left\{\delta_{2}(1)A_{1i}+\delta_{2}(2)A_{2i}\right\}R_{i} -\gamma A_{2i}(1-R_{i})\\
&&.\\
&&.\\
&&.\\
Y_{Ki}&-&\sum^{K}_{j=1}\delta_{K}(j)A_{ji}R_{i}-\gamma A_{Ki}(1-R_{i})
\end{pmatrix}\nonumber
\label{eq:Score2}
\end{eqnarray}

\subsection{Further technical details on model fitting and inference}
\label{sec:technical}

\subsubsection{Variance estimation for $\hat{\theta}$}

After minimising a given score equation to obtain $\hat{\theta}$, its variance-covariance matrix can be approximated by the sandwich expression \cite{newey1994}

\begin{equation}
Var(\hat{\theta}) =  \frac{1}{n}G^{-1}_{\text{Gen}}(\hat{\theta})\text{Var}[S_{i}(\hat{\theta})]G^{-1}_{\text{Gen}}(\hat{\theta})^{'} \label{eq:sandwich1}
\end{equation}

\noindent The middle term is estimated by taking the sample variance of the individual score terms at $\theta=\hat{\theta}$, which is a $K\times K$ matrix. To calculate the outer term we execute the following procedure:

\begin{enumerate}
 \item{Calculate the gradient matrix $\left(\frac{\partial S_{i}(\theta)}{\partial\theta}\right)$ for each subject - whose $jl^{th}$ element is the derivative of the $j^{th}$ component of $S_{i}(\theta)$ with respect to the $l^{th}$ element of $\theta$. Note that this is not a square $K\times K$ square matrix, but a $K\times 2$ or $K\times 3$ matrix for our two and three parameter models respectively;}
    \item{Evaluating the result of step 1 at $\hat{\theta}$;}
    \item{Taking the sample average for the result of step 2, so that $G(\hat{\theta}) = E\left(\frac{\partial S_{i}(\theta)}{\partial \theta}\rvert_{\theta=\hat{\theta}}\right)$;}
    \item{Calculating the generalized inverse $G^{-1}_{\text{Gen}}(\hat{\theta}) = \left\{G^{'}(\hat{\theta}) G(\hat{\theta})\right\}^{-1}G^{'}(\hat{\theta})$}
\end{enumerate} 

\noindent In the Appendix, we perform a simulation study to verify that the G-estimation strategies outlined in Section 2.3 and 2.5 can return unbiased estimates for the model parameters furnishing Hypothetical Estimand 1 and 2 when the structural mean model is correctly specified. Moreover, we show that standard errors obtained from sandwich variance formula (\ref{eq:sandwich1}) also accurately quantify the true uncertainty in parameter estimates of the structural mean model. R code is provided for reproducibility.

\subsubsection{Incorporating covariates into the model}
\label{sec:covariates}

In order to improve the efficiency of the G-estimation approach, we make use of a full set of $v$ baseline covariates $\boldsymbol{C}$ when solving the system of score equations. Taking, for example, Hypothetical Estimand 2, we assume that
\begin{eqnarray}
E\left\{Y_{k}|R=1,\boldsymbol{C},\bar{A}_{k}\right\}-E\left\{Y_{k}(1,\boldsymbol{0}_{k})|R=1,\boldsymbol{C},\bar{A}_{k}\right\} &=& \sum^{k}_{j=1}\delta_{k}(j)A_{j},
\nonumber \\
E\left\{Y_{k}|R=0,\boldsymbol{C},\bar{A}_{k}\right\}-E\left\{Y_{k}(0,\boldsymbol{0}_{k})|R=0,\boldsymbol{C},\bar{A}_{k}\right\} &=& \sum^{k}_{j=1}\gamma_{k}(j)A_{j}, \nonumber
\end{eqnarray}

This approach assumes no effect modification by $\boldsymbol{C}$. More efficient estimators may then be obtained by using the earlier estimation strategy upon replacing each adherence and weight outcome at time point $k$ by its residual, once the covariate effects have been partialed out. That is we replace: 
\begin{itemize}
    \item{$Y_{k}$ with $Y^{C}_{k} = Y_{k}-E(Y_{k}|\boldsymbol{C})$;}
    \item{$A_{k}$  with $A^{C}_{k} = A_{k}-E(A_{k}|\boldsymbol{C},R=1)$ if $R=1$ and $A^{C}_{k} = A_{k}-E(A_{k}|\boldsymbol{C},R=0)$ if $R=0$;}
\end{itemize}
\noindent where $E(.|\boldsymbol{C})$ are obtained as fitted values from a user specified regression model. Fortuitously, due to the independence of $R$ and $\boldsymbol{C}$, the uncertainty in these fitted values can be ignored in the variance calculation \cite{newey1994}. 

\subsubsection{Schema for full uncertainty quantification of weight loss trajectories and Hypothetical Estimands in the STEP 1 data}
\label{sec:MI}

Fully analytic expressions for the uncertainty in the Hypothetical Estimands are difficult to obtain for two reasons: Firstly, complete weight trajectory data was missing for $\approx$ 17\% of participants in STEP 1. Secondly, the weight loss trajectories used to calculate the Hypothetical Estimands are complex non-linear functions of the estimated model parameters, which are more complicated to approximate via a Taylor series. To address these issues we applied the following uncertainty quantification schema incorporating multiple imputation, analytic sandwich variance approximation and Monte-Carlo simulation steps: \\
\\
We used multiple imputation to create a series of complete data sets $D^{*}_{q}, q=1\ldots,m$, each of dimension $n\times(2K+v+1)$, where: 
\begin{equation}
D^{*}_{q} =  (\boldsymbol{Y}^{*}_{q1},\ldots, \boldsymbol{Y}^{*}_{qK},\boldsymbol{\bar{A}_{K}},\boldsymbol{C},\boldsymbol{R}),  \nonumber  
\end{equation} 
\noindent Here the superscript $^{*}$ (and lack thereof) denotes variables that vary across imputed data sets. Imputation was performed in four separate sub-groups:
\begin{itemize}
    \item Treatment arm full adherers ($R=1$ \& $\bar{A}_{K} = \boldsymbol{1}_{K}$);
    \item Treatment arm non-full adherers ($R=1$ \& $\bar{A}_{K} \neq \boldsymbol{1}_{K}$);
    \item Placebo arm full adherers ($R=0$ \& $\bar{A}_{K} = \boldsymbol{1}_{K}$);
    \item Placebo arm non-full adherers ($R=0$ \& $\bar{A}_{K} \neq \boldsymbol{1}_{K}$)
\end{itemize}
We used multiple imputation via Chained Equations (MICE) software \cite{white2011} (implemented via predictive mean matching) for this purpose.  For each complete data set $D^{*}_{q}$ we calculated $\hat{\theta}^{*}_{q}$ by solving the relevant score equation and from that its sandwich variance estimate $\Sigma^{*}_{q}$. We then used the Rubin rule \cite{rubin2018} to calculate pooled point estimates $\hat{\theta}$ and variances  $Var(\hat{\theta})$, and the entire covariance matrix $\hat{\Sigma}$ across the $m$ imputed data sets.\\
\\
To calculate point estimates and confidence intervals for Hypothetical Estimand 1 we sampled $N$=10,000 parameter values $\hat{\theta}_{1},\ldots,\hat{\theta}_{N}$ from a MVN$_{2}(\hat{\theta},\hat{\Sigma})$ distribution. For each $N$ parameter draws we constructed the full weight loss trajectory implied by parametric model (\ref{eq:decaymodel1}) at time points $k$=1,...,$K$. From this we calculated their time-point specific sample mean and variance. The same procedure was applied for Hypothetical Estimand 2, using the relevant MVN$_{3}(\hat{\theta},\hat{\Sigma})$ distribution to draw weight loss trajectories from model (\ref{eq:decaymodel2}).




\section{Application to the STEP 1 trial data}

We now apply our proposed structural mean models to adjust for non-adherence in the treatment arm only (Hypothetical Estimand 1), or in both arms (Hypothetical Estimand 2), using data from the STEP 1 trial. Of the 1961  individuals within STEP 1, 1629 had complete information on their weight outcome history at all 12 time points available (i.e. week's 2, 4, 8, 12, 16, 20, 28, 36, 44, 52, 60 and 68), so that $t_{1}=2$, $t_{2}=4$, and so on. For model fitting, Multiple Imputation using Chained Equations (MICE) \cite{white2011} was firstly used to create $m=5$ data sets via predictive mean matching, making use of a covariate vector $\boldsymbol{C}$ consisting of baseline weight, sex, age, HbA1c and country of study. Out of the 16 countries contributing participants to STEP 1, the largest contingent (38\%) were from the USA and the second largest (11\%) were from the UK. Weight outcome data was modelled on the percent change scale, in line with previous publications. To assess the benefit of covariate adjustment in the outcome model and adherence variables (see Section \ref{sec:covariates}), structural mean models were implemented after adjustment for the full covariate set $\boldsymbol{C}$ and using baseline weight, $Y_{0}$, alone. In each case, this was implemented using simple linear regression models for $E(.|\boldsymbol{C})$. \\
\\
When minimising the relevant score equation $S(\theta)$ to furnish the parameters of Hypothetical Estimand 1 and 2, we used the identity matrix in place of $\boldsymbol{\Sigma}$. Estimates were calculated separately for each fully imputed data set and then combined using the Rubin rule, as described in Section \ref{sec:MI} \cite{wilding2021}. Full results for all analyses are shown in Table \ref{tab:basicDecayModel}. Parameter and Hypothetical Estimand estimates are shown to two decimal places, except those for the decay parameter $\alpha$ which are specified to four decimal places - its precise value above or below one warrants additional accuracy due to its subsequent impact on the time-varying causal effect predictions.  

\subsection{Adherence adjustment in treatment arm only}

\begin{figure}[!hbtp]
    \centering
    \includegraphics[width=0.49\textwidth,clip]{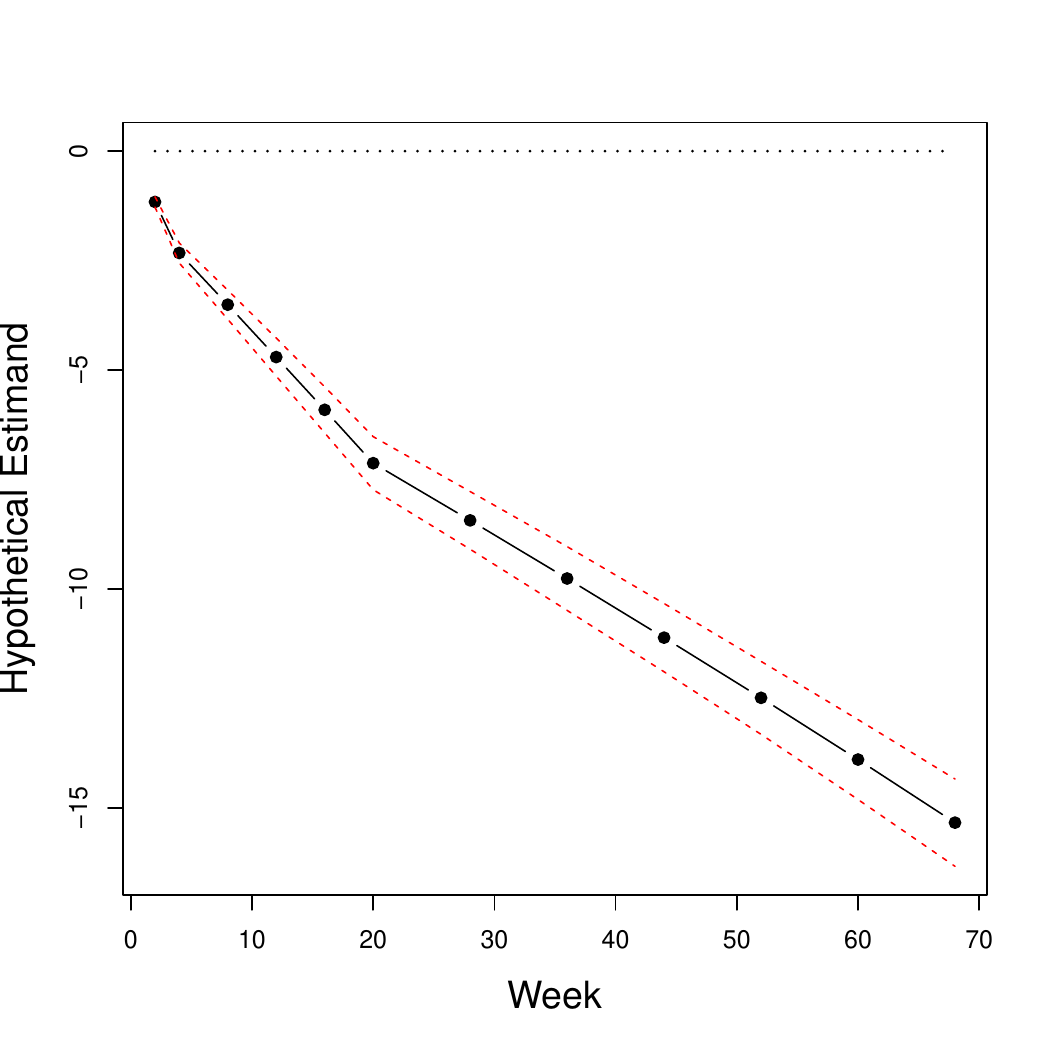}
    \includegraphics[width=0.49\textwidth,clip]{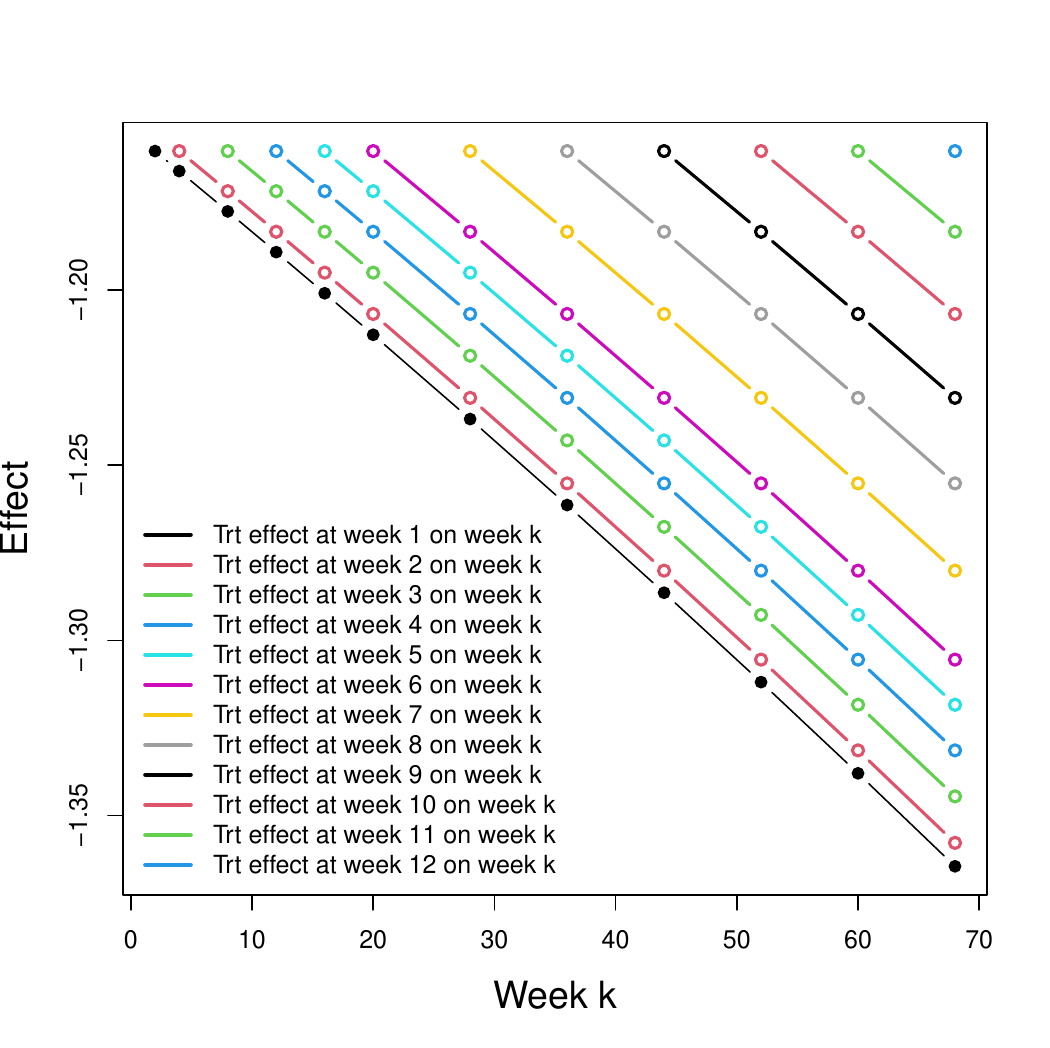}    
    \caption{\textit{Left: Estimated average causal effects of treatment administered at each time point on weight at subsequent future time points obtained from Model 1. Right: Hypothetical Estimand 1 estimates and confidence intervals for the effect of receiving treatment in full versus none up to week $k$, for $k\in \{2,...,68\}$. The week 68 point estimate and 95\% confidence interval corresponds to Hypothetical Estimand 1.}}
    \label{fig:basicDecayModel}
\end{figure}

When modelling non-adherence in the treatment arm only, and incorporating the full covariate vector $\boldsymbol{C}$ into the analysis, structural model (\ref{eq:decaymodel1}) estimates an immediate effect of treatment at time point $k$ on weight at week $k$ as a reduction of 1.17\%. Surprisingly, we see that the decay parameter $\alpha$ governing the effect of earlier treatment on later weight is slightly larger than 1. Figure \ref{fig:basicDecayModel} (left) shows that treatment taken at week 2 is inferred to induce a 1.35\% weight reduction  at week 68  (bottom-right black dot). This interpretation is arguably biologically implausible. Nevertheless, figure \ref{fig:basicDecayModel} (right) shows the point estimate and confidence intervals for the weight loss trajectory under full treatment adherence implied by model (\ref{eq:decaymodel1}) at each of the 12 time points in the trial. Hypothetical Estimand 1, the mean difference in percentage weight loss under full adherence to treatment at week 68, is estimated to be a 15.25\% reduction (95\% CI: -14.43\%,-16.07\%). In this case, full covariate adjustment is seen to produce near identical point estimates and confidence intervals compared to using baseline weight alone (Table \ref{tab:basicDecayModel}).

\subsection{Adherence adjustment in treatment and placebo arms}

\begin{figure}[hbtp]
    \centering
    \includegraphics[width=0.49\textwidth,clip]{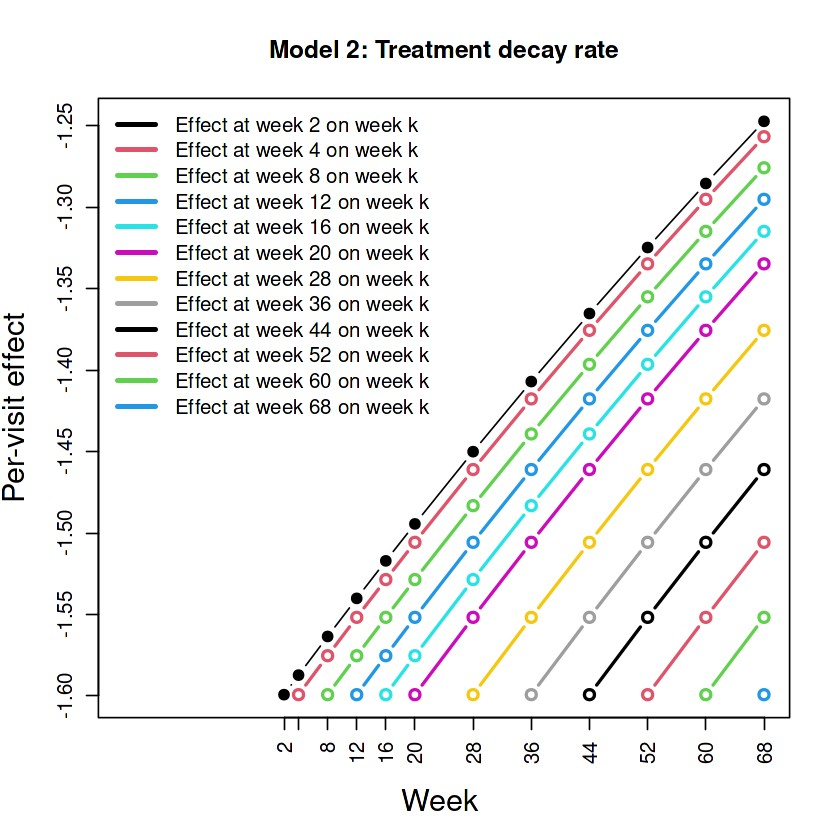}
    \includegraphics[width=0.49\textwidth,clip]{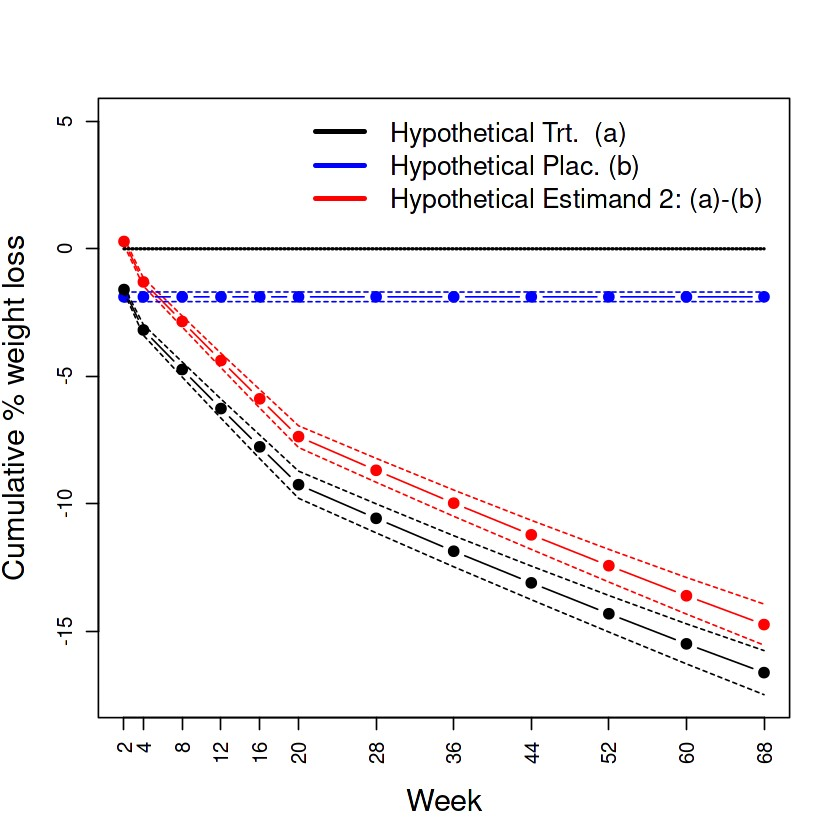}
    \caption{\textit{Left: Estimated average causal effects of treatment administered at each time point on weight at subsequent future time points under Model 2. Right: Estimand Hypothetical Estimands and confidence intervals for the effect of receiving (a) treatment in full versus none and (b) placebo in full versus none up to week $k$, for $k\in \{2,...,68\}$. The difference between the treatment and placebo trajectories is shown in red, with  The week 68 point estimate and  95\% confidence interval of the red curve corresponds to Hypothetical Estimand 2.}}
    \label{fig:basicDecayModel2}
\end{figure}

Table \ref{tab:basicDecayModel} and Figure \ref{fig:basicDecayModel2} show the results of fitting structural causal model (\ref{eq:decaymodel2}) accounting for non-adherence in both the treatment and placebo arms. Under this model, and using full covariate adjustment, the causal effect of taking treatment at week $k$ is a (larger) 1.6\% reduction in weight at week $k$. To counter-balance this, the decay parameter $\alpha$ is estimated to be less than 1, meaning that effect of taking treatment at week $k$ on weight in future weeks follows a more plausible diminishing trend (Figure  \ref{fig:basicDecayModel2} left). For example, the total weight loss at week 68 that is attributable to treatment at week 2 is $\approx$ -1.25\%. In the placebo arm, the constant instantaneous placebo effect parameter $\gamma$ is estimated to be -1.89\%.\\
\\
Figure  \ref{fig:basicDecayModel2} (right) shows three implied weight loss trajectories: Firstly, the weight change in the treatment arm under full adherence versus no adherence (black line, contrast (a) in Table \ref{tab:basicDecayModel}; secondly, the weight change in the placebo arm under full adherence versus no adherence (blue line, contrast (b) in Table \ref{tab:basicDecayModel}); and thirdly, the difference between (a) and (b) (red line), whose 12th and final time-point at week 68 is an estimate for Hypothetical Estimand 2. At week 68 our model implies that there would have been a 16.63\% weight reduction in the treatment arm, leading to an estimate for Hypothetical Estimand 2 of -14.75\%, 95\% C.I. (-13.94\%,-15.55\%). In this case, full covariate adjustment is seen to produce similar point estimates and confidence intervals for Hypothetical Estimand 2 compared to using baseline weight alone (Table \ref{tab:basicDecayModel}). However,  standard errors for $\hat{\beta},\hat{\alpha}$ and $\hat{\gamma}$ are noticeably smaller when all covariates are used.

\begin{table}[ht]
\centering
\begin{tabular}{rrrrrr}
  \hline
Parameter/Estimand & Estimate & S.E & Lower C.I & Upper C.I & p-value \\ 
  \hline
  \multicolumn{6}{c}{{}}\\
\multicolumn{6}{c}{{\bf Adherence adjustment in treatment arm only: $\theta=(\beta,\alpha)$ ($\boldsymbol{C}$=$Y_{0}$)}}\\
$\beta$ & -1.17 & 0.04 & -1.24 & -1.09 & $<$1$\times$10$^{-16}$ \\ 
$\alpha$ & 1.0021 & 0.0008 & 1.0005 & 1.0036 & 0.0084 \\
  Hypothetical Estimand 1 & -15.17 & 0.42 & -16.00 & -14.34 & $<$1$\times$10$^{-16}$ \\
\multicolumn{6}{c}{{}}\\
\multicolumn{6}{c}{{\bf Adherence adjustment in both arms: $\theta=(\beta,\alpha,\gamma)$ ($\boldsymbol{C}$=$Y_{0}$)}}\\  
$\beta$ & -1.63 & 0.07 & -1.76 & -1.49 & $<$1$\times$10$^{-16}$ \\ 
  $\alpha$ & 0.9958 & 0.0009 & 0.9940 & 0.9976 & 6$\times$10$^{-6}$ \\ 
  $\gamma$ & -2.00 & 0.19 & -2.36 & -1.63 & $<$1$\times$10$^{-16}$ \\ 
 Treatment arm contrast (a) & -16.62 & 0.45 & -17.51 & -15.74 & $<$1$\times$10$^{-16}$ \\ 
  Placebo arm contrast (b) & -1.99 & 0.19 & -2.36 & -1.63 & $<$1$\times$10$^{-16}$ \\ 
  Hypothetical Estimand 2: (a)-(b) & -14.63 & 0.41 & -15.44 & -13.82 & $<$1$\times$10$^{-16}$ \\ 
\multicolumn{6}{c}{{}}\\
\multicolumn{6}{c}{{\bf Adherence adjustment in treatment arm only: $\theta=(\beta,\alpha)$ ($\boldsymbol{C}$=Full set)}}\\  
$\beta$ & -1.17 & 0.04 & -1.25 & -1.09 & $<$1$\times$10$^{-16}$ \\ 
  $\alpha$ & 1.0022 & 0.0008 & 1.0007 & 1.0037 & 0.0049 \\ 
  Hypothetical Estimand 1: & -15.25 & 0.42 & -16.09 & -14.42 & $<$1$\times$10$^{-16}$ \\
\multicolumn{6}{c}{{}}\\
\multicolumn{6}{c}{{\bf Adherence adjustment in both arms: $\theta=(\beta,\alpha,\gamma)$ ($\boldsymbol{C}$=Full set)}}\\  
$\beta$ & -1.60 & 0.056 & -1.71 & -1.49 & $<$1$\times$10$^{-16}$ \\ 
  $\alpha$ & 0.9962 & 0.0008 & 0.9947 & 0.9978 & 3$\times$10$^{-6}$ \\ 
  $\gamma$ & -1.89 & 0.10 & -2.09 & -1.68 & $<$1$\times$10$^{-16}$ \\ 
 Treatment arm contrast (a) & -16.63 & 0.44 & -17.49 & -15.77 & $<$1$\times$10$^{-16}$ \\ 
Placebo arm contrast (b) & -1.88 & 0.10 & -2.08 & -1.68 & $<$1$\times$10$^{-16}$ \\ 
 Hypothetical Estimand 2: (a)-(b) & -14.75 & 0.41 & -15.55 & -13.94 & $<$1$\times$10$^{-16}$ \\ 
   \hline
\end{tabular}
\caption{\textit{Point estimates, standard errors, confidence intervals and p-values for the structural mean model parameters $(\beta,\alpha,\gamma)$ and the implied Hypothetical Estimands. Results are shown on the percentage reduction scale}}
\label{tab:basicDecayModel}
\end{table}

\section{Discussion}

In this paper we  propose a structural mean modelling framework for targeting hypothetical estimands in the landmark STEP 1 trial. It uses randomization as an Instrumental Variable and is able to fully exploit longitudinal data on the weight loss outcome. As such, it represents a significant extension to previously proposed IV methods within the Estimand Framework \cite{bowden2021}. Our decay model formulation developed specifically for STEP 1 is nevertheless well suited to modelling general outcome trajectories, whose complex dynamics are affected by an individual's complete adherence history. Extensions of the methodology beyond a continuous outcome to the binary or time-to-event setting will be pursued as further work. We view it as an attractive alternative to the approach of Olarte Parra et. al \cite{parra2022}, which can incorporate time-varying covariates (including previous adherence and outcome measures) into estimators based on inverse probability weighting or G-computation, due to not relying on the sequential ignorability assumption nor being sensitive to positivity violations. In a broader sense, lack of sensitivity to positivity violation is important because clinical trials can face strongly selective intercurrent events (e.g., treatment discontinuation, rescue treatments) which may be (near-)deterministically indicated when severe adverse events, such as disease progression, occur. \\
\\
The suggested robustness of our IV approach comes at the expense of modelling assumptions on the functional form of the causal effect and who it applies to: specifically equations (\ref{eq:homogeneityT}) and (\ref{eq:decaymodel1}) for Hypothetical Estimand 1, and
equations (\ref{eq:homogeneityT}), (\ref{eq:homogeneityP}) and (\ref{eq:decaymodel2}) for Hypothetical Estimand 2. We view these assumptions to be weaker than assuming that measurements are available on all prognostic factors of the outcome that influence the decision to discontinue treatment, and that the association of these measurements with treatment discontinuation or the outcome has been correctly modelled. Furthermore, whereas our modelling assumptions are guaranteed to hold under the null hypothesis of no treatment effect, meaning that the proposed approach is guaranteed to deliver robust tests of that null hypothesis, this is not the case for approaches that rely on sequential ignorability.\\
\\
Our focus in this study was on Hypothetical Estimands, which attempt to quantify effects of interventions on the full trial population. An alternative but closely related strategy mentioned in ICH E9 guidance \cite{ICHE9} is to target a Principal Stratum estimand. This general strategy focuses on the  effect of (randomized) treatment assignment in a specific participant subgroup defined by their potential adherence to one or more treatments in the trial \cite{bowden2021}. For example, Qu et al. (2020)\cite{qu2020general} propose two Principal Strata of interest when conducting a general analysis within the Estimand framework: firstly, participants who would fully adhere to randomized treatment in both arms; and secondly, participants who would fully adhere to the active treatment if randomized to receive it, but who may or may not adhere to placebo if randomized to receive it. Strong assumptions are needed to learn the effect for these patient strata, because the observed data do not directly enable one to learn {\it which} participants would be adherent to active treatment as well as placebo.  Qu et al \cite{qu2020general} and others \cite{Bornkamp2020,lu2022} identify these strata by assuming the availability of post-treatment measurements $Z$ (e.g., safety outcomes) with counterfactual values $Z(R=0)$ and $Z(R=1)$ that are conditionally independent given baseline covariates. Future work will focus on hypothetical estimands for observable participant subgroups defined at an intermediate time (e.g., week 20 treatment `responders' in the context of STEP 1). We believe these can be identified with minor adaptations of the causal model considered here, thereby avoiding such strong assumptions.\\
\\
In estimating equations (\ref{eq:ScoreNaive}) and (\ref{eq:Score2}) , we used the identity matrix for $\boldsymbol{\Sigma}$, but a more complex choice that can in theory deliver better efficiency is the residual covariance matrix of the predicted adherence-free potential outcomes. This quantity depends on the unknown model parameters and must therefore be updated at each iteration of the score function minimisation. An advantage of this is that the score function then follows a known $\chi^{2}$ distribution. As further work, we plan to incorporate this into our software, as well as developing a framework to assess the goodness of fit of a range possible causal models, with a view to further improving the performance and power of our method. \\

\appendix


\section*{Appendix 1: Monte Carlo simulations}

We perform Monte Carlo simulations to sanity check the validity of our proposed G-estimation method and sandwich variance formula (\ref{eq:sandwich1}) for quantifying Hypothetical Estimand 1 and 2. To mimic the setup of STEP 1, we generate simulated data sets with $n = 1961$ individuals across $K = 12$ time points. For simplicity, we make the time intervals equally spaced, so that $t_{k} = k$, $k=1,..., 12$. For each individual $i$, their initial treatment assignment $R_i$ is generated from a Bernoulli distribution with $Pr(R_i = 1)=\frac{1}{2}$, with $R_{i}$=1 being the experimental treatment and $R_{i}$=0 being placebo. For individual $i$ at time point $k$, we denote their adherence status by $A_{i, k}$, whereby $A_{i, k} = 1$ indicates that they take the real (or sham in case of placebo) injection as intended, and $A_{i, k} = 0$ if they do not. We can then write the actual adherence status for individual $i$ at time point $k$ as $A_{i, k}R_i$ in the treatment arm or $A_{i, k}\left(1 - R_i \right)$ in the placebo arm. 

\subsection*{Adherence status generation}

Let $PT_{i, k}$ ($PP_{i, k}$) be the probability of $A_{i, k} = 1$ for individuals in the treatment (placebo) arm. $PT_{i, k}$ is determined by the following logit model:
\begin{equation*}
    PT_{i, k} = \frac{e^{(\eta_0 + \eta_1 A_{i, k-1} + \eta_2 Y_{i, k-1} + \eta_k k + U_{i, k})}}{1 + e^{(\eta_0 + \eta_1 A_{i, k-1} + \eta_2 Y_{i, k-1} + \eta_k k + U_{i, k})}},
\end{equation*}
where
\begin{itemize}
    \item{$A_{i, k-1}$ and $Y_{i, k-1}$ are the adherence status and outcome at the previous time point;}
    \item{$U_{i, k}$ is an unobserved serially correlated confounder generated by the following auto-regressive AR(1) process:
\begin{equation*}
    U_{i, k} = 0.98 U_{i, k-1} + N(0. 0.2^2),
\end{equation*}
where $U_{i, k-1} = 0$ with $k = 1$;}
\item{$\eta_{0}\ldots\eta_{k}$ are fixed parameters.}
\end{itemize}
When $k = 1$, we set $A_{i, k-1} = Y_{i, k-1} = 0$ and ignore the time effect by setting $\eta_k = 0$.  We set $\eta_0 = 3$, $\eta_1 = 0.2$, $\eta_2 = -0.1$ and $\eta_k = -0.2$. These coefficients imply that an individual is more likely to adhere if they have previously and if they experience a larger weight loss. It also induces a decrease in the probability of adherence over time. Similarly, we generate adherence status variables $A_{i, k}$ in the placebo arm from a Bernoulli distribution with
\begin{equation*}
    PP_{i, k} = \frac{e^{(\pi_0 + \pi_1 A_{i, k-1} + \pi_2 Y_{i, k-1} + \pi_k k + U_{i, k})}}{1 + e^{(\pi_0 + \pi_1 A_{i, k-1} + \pi_2 Y_{i, k-1} + \pi_k k + U_{i, k})}},
\end{equation*}
where $\pi_0 = 3$, $\pi_1 = 0.3$, $\pi_2 = -0.25$ and $\pi_k = -0.2$.

\subsection*{Outcome generation}

\noindent For transparent estimation of the parameters furnishing Hypothetical Estimand 1, we first generate outcome $Y_{i, k}$  from the following model, assuming the placebo effect is zero:
\begin{equation}\label{eq:simulation_E1}
    Y_{i, k} = \sum_{t = 1}^k \beta \alpha^{k - t} A_{i, t}R_i + U_{i, k}.
\end{equation}
For transparent estimation of the parameters furnishing Hypothetical Estimand 2, we generate $Y_{i, k}$ from the following model which incorporates a non-zero placebo effect:
\begin{equation}\label{eq:simulation_E2}
    Y_{i, k} = \sum_{t = 1}^k \beta \alpha^{k - t} A_{i, t}R_i + \gamma A_{i, k}(1 - R_i) + U_{i, k},
\end{equation}
where $\gamma$ is set to -0.9. For simplicity, we do not include baseline covariates in the adherence or outcome models. They can, however,  be viewed as residuals after partialling  out baseline covariates (see Section 2.6.2). \\
\\
The proportion of non-adherers within each trial arm and at each of the 12 time points, is shown for a single representative simulated dataset in Figure~\ref{fig:adherence_simulation} below.
\begin{figure}[H]
    \centering
    \includegraphics[width=0.7\textwidth]{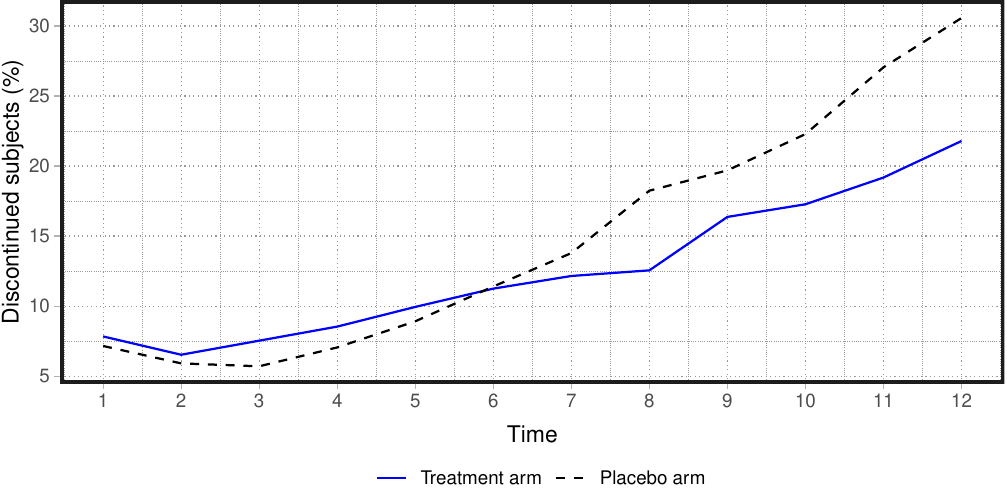}
    \caption{{\it Adherence rates in the treatment and placebo arm over the 12 time periods of the trial for a representative single data set.}}
    \label{fig:adherence_simulation}
\end{figure}
\noindent Over 1000 simulated data sets, we estimated causal parameters $\beta$ and $\alpha$ using the G-estimation method elaborated in Section \ref{sec:EstSMM1} with outcome data generated by Equation~(\ref{eq:simulation_E1}). Using 1000 simulated data sets from outcome model ~(\ref{eq:simulation_E2}), we then estimated $\beta$, $\alpha$ and $\gamma$ using the the G-estimation strategy in Section \ref{sec:EstSMM2} for targeting Hypothetical Estimand 2. In both cases we set $\Sigma$ as the $12 \times 12$ identity matrix. The results are presented below in Table~\ref{tab:simulation_results}. We report three summary statistics across the 1000 simulations: ``\textit{Estimate}'': the mean values of the parameter estimates; ``\textit{Empirical SE}'': the standard deviation of the parameter estimates; ``\textit{Sandwich SE}'' is the mean standard error of the estimates calculated from the sandwich variance formula  in Section 2.6.1. For both models the G-estimation method works as intended, with mean point estimates close to the ground truth, and empirical standard errors matching those obtained from the sandwich variance formula. 

\begin{table}[hbtp]
\centering
\begin{tabular}{lccc}
\hline
 & $\beta$ & $\alpha$ & $\gamma$ \\ 
 \hline
\textbf{Truth}     & $-1.1$ & $0.95$ & $-0.9$ \\ 
\hline
\multicolumn{4}{l}{\textbf{Estimand 1:}} \\ 
\hline
Estimate     & -1.102 & 0.949 & - \\
Empirical SE & 0.017 & 0.003 & - \\
Sandwich SE  & 0.017 & 0.002 & - \\ 
\hline
\multicolumn{4}{l}{\textbf{Estimand 2:}} \\ 
\hline
Estimate     & -1.103 & 0.950 & -0.907 \\
Empirical SE & 0.009 & 0.001 & 0.051 \\
Sandwich SE  & 0.009 & 0.001 & 0.052 \\ 
\hline
\end{tabular}
\caption{{\it Simulation results: mean point estimates, empirical standard errors and mean sandwich variance standard errors of the causal parameters furnishing Hypothetical Estimand 1 and 2, over $1000$ replications.}}
\label{tab:simulation_results}
\end{table}

\noindent R code to reproduce the simulation study is available in {\it Online Supplementary Material}.

\appendix

\appendix

\end{document}